\documentclass[3p,times,procedia]{elsarticle}
\usepackage{nupha_ecrc}

\newcommand{\pp}{\ensuremath{\mathrm {p\kern-0.05em p}}}

\newcommand{\sqrtSnn}{\ensuremath{\sqrt{s_{\mathrm{NN}}}}}

\newcommand{\pt}{\ensuremath{p_{\mathrm{T}}}}
\newcommand{\ptch}{\ensuremath{p_{\mathrm{T, ch jet}}}}

\newcommand{\MeVc}{\ensuremath{\mathrm{MeV}\kern-0.05em/\kern-0.02em c}}
\newcommand{\GeVc}{\ensuremath{\mathrm{GeV}\kern-0.05em/\kern-0.02em c}}
\newcommand{\GeVcSq}{\ensuremath{\mathrm{GeV}\kern-0.05em/\kern-0.02em c^2}}

\newcommand{\jpsi}{\ensuremath{{\rm J}\kern-0.02em/\kern-0.05em\psi}}

\newcommand{\ptjet}{\ensuremath{p_{\mathrm{T}}^{\rm ch~jet}}}

\newcommand{\ptraw}{\ensuremath{\ptch^{\rm raw}}}

\newcommand{\kt}{\ensuremath{k_{\mathrm{T}}}}
\newcommand{\vj}{\ensuremath{v_{2}^{\mathrm{ch~jet}}}}

\newcommand{\vjfull}{\ensuremath{v_{2}^{\mathrm{ch+em jet}}}}

\newcommand{\vpart}{\ensuremath{v_{2}^{\mathrm{part}}}}
\newcommand{\dpt}{\ensuremath{\delta \kern-0.15em p_{\mathrm{T}}}}
\newcommand{\rholocal}{\ensuremath{\rho_{\mathrm{ch~local}}}}
\newcommand{\rhovar}{\ensuremath{\rho_{\mathrm{ch}}(\varphi)}}
\newcommand{\rhoav}{\ensuremath{\langle \rho_{\rm ch} \rangle}}

\newcommand{\ep}{\ensuremath{\Psi_{\mathrm{EP,~2}}}}
\newcommand{\epthree}{\ensuremath{\Psi_{\mathrm{EP,~3}}}}

\newcommand{\epn}{\ensuremath{\Psi_{\mathrm{EP},~n}}}

\newcommand{\rpn}{\ensuremath{\Psi_n}}
\newcommand{\rptwo}{\ensuremath{\Psi_2}}
\newcommand{\phijet}{\ensuremath{\varphi_{\mathrm{jet}}}}

\newcommand{\nin}{\ensuremath{N_{\mathrm{in}}}}
\newcommand{\nout}{\ensuremath{N_{\mathrm{out}}}}

\volume{00}

\firstpage{1}

\journalname{Nuclear Physics A}

\runauth{R. A. Bertens}

\jid{nupha}

\jnltitlelogo{Nuclear Physics A}

\usepackage{amssymb}

\usepackage[figuresright]{rotating}

\begin{document}

\begin{frontmatter}


\title{Azimuthal anisotropy of charged jet production \\ in $\mathbf{\sqrt{{\textit s}_{\rm NN}}}$~=~2.76~TeV~Pb--Pb collisions}
\author{R.A. Bertens (for the ALICE collaboration)}
 \ead{redmer.alexander.bertens@cern.ch}
 \address{Utrecht University, P.O.Box 80000, 3508 TA Utrecht, The Netherlands}

\dochead{Quark Matter 2015}


\begin{abstract}
    Measurements of charged jet production with respect to the second-order harmonic event plane, quantified as \vj{}, in \sqrtSnn{} = 2.76 TeV Pb--Pb collisions are presented. The contribution of hydrodynamic flow to the underlying event energy is corrected for on a jet-by-jet basis. Remaining effects from instrumental resolution and regional fluctuations in the energy density of the underlying event are corrected for by unfolding. Significant \vj{} is observed in semi-central collisions. Qualitative agreement is found with \vjfull{} of jets reconstructed from charged and neutral fragments and $v_2$ of single charged particles, as well as with predictions from the JEWEL Monte Carlo model.
\end{abstract}


\end{frontmatter}

\section{Introduction}
Jets in heavy-ion collisions are used to probe the quark-gluon plasma (QGP), as medium-induced parton energy loss from elastic and radiative interactions between partons and the QGP lead to a modification of the measured jet spectrum \cite{gluonradiation,gluonradiation2}. The dependence of the energy loss on the in-medium path-length provides insight into the energy-loss mechanisms and can be studied by measuring jet production relative to the orientation of the second-order symmetry plane \rptwo{}. The azimuthal asymmetry in the jet production is quantified as \vj{}, the second-order coefficient of the Fourier expansion of the azimuthal distribution of jets relative to \rpn,
\begin{equation}\label{eq:flowdud}
    \frac{\mathrm{d}N}{\mathrm{d}\left(\phijet - \rpn\right)} \propto 1 + \sum_{n=1}^{\infty} 2 v_n^{\mathrm{jet}} \cos\left[n\left(\phijet - \rpn\right)\right],
    \end{equation}
    where \phijet{} denotes the azimuthal angle of the jet and \rpn{} are the orientations of the symmetry axes of the initial nucleon distribution of the collision overlap region. 
    
    These proceedings give an overview of the new ALICE results of \vj{} measurements in central and Pb--Pb semi-central collisions, as well as model comparisons and comparisons to other observables which are sensitive as well to in-medium parton energy loss.  

\section{Analysis and experimental setup}
The data used in this measurement were recorded in 2010 and 2011 with the ALICE detector \cite{review} and comprise 6.8$\times$10$^6$ central collisions (0-5\% collision centrality) and 8.6$\times$10$^6$ semi-central collisions (30-50\%) with a primary vertex within $\pm$10cm of the nominal interaction point. Charged particle tracks are reconstructed by the Inner Tracking System and Time Projection Chamber at $\vert \eta \vert <$ 0.9, employing track selection criteria which optimize momentum resolution while retaining full azimuthal acceptance. Centrality determination, as well as reconstruction of the event planes \epn{} (the experimental estimates of \rpn{}, reconstructed from the density of particles emitted in the transverse plane) is performed by the V0 detectors, located at $2.8 < \eta < 5.1$ and  $-3.7 < \eta < -1.7$. 

\begin{figure}
    \begin{center}
        \includegraphics[width=0.5\textwidth]{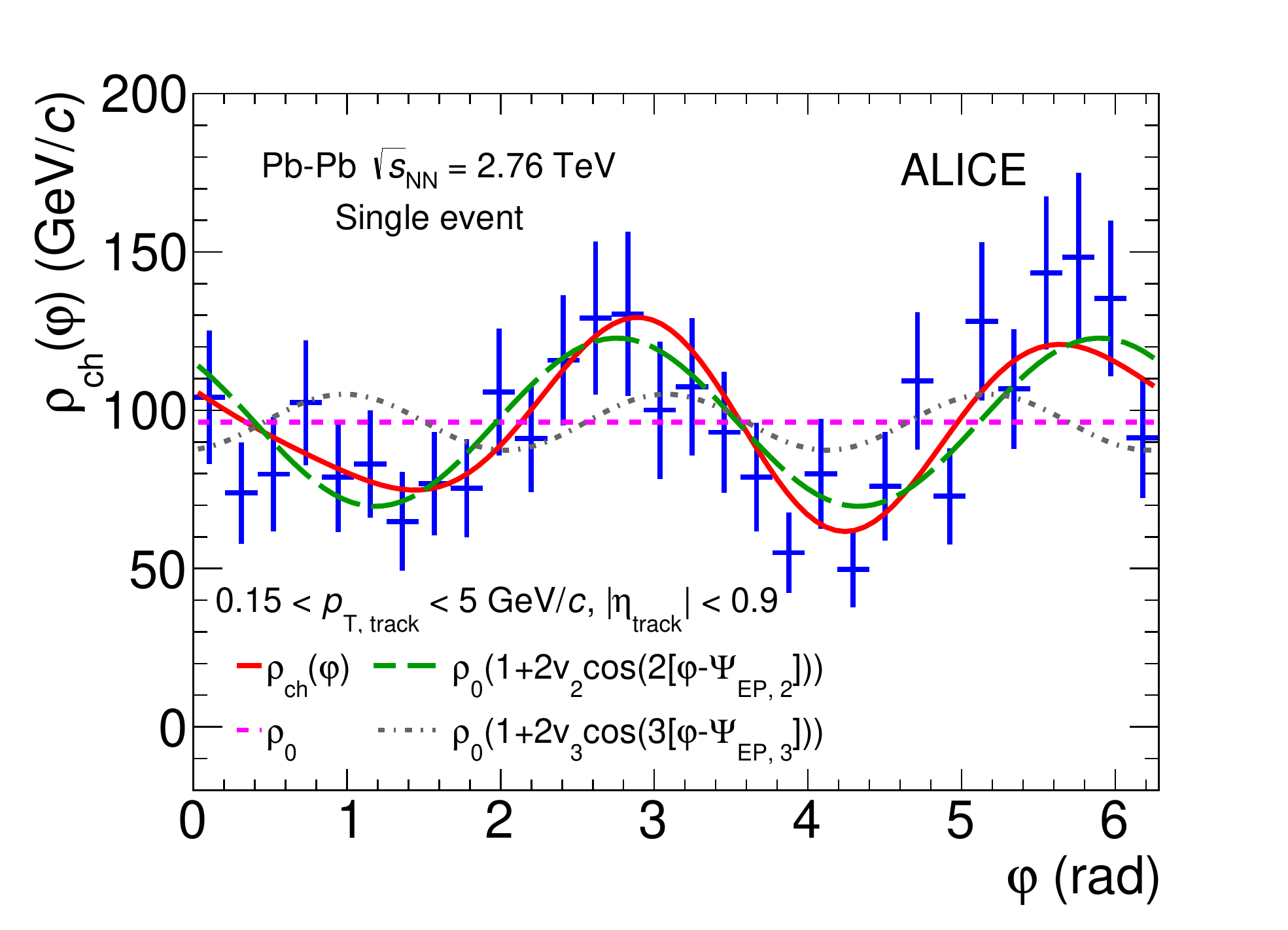}
        \caption{Single event example of charged track transverse momentum density as a function of azimuthal angle $\varphi$. The red curve represents the fit of Eq.~\ref{eq:master} to the data. Independent contributions of $v_2$ and $v_3$, obtained from Eq.~\ref{eq:master} as well, are shown as the green and gray curves. The normalization constant $\rho_0$ is given by the dashed magenta line. Event from the 0--5\% centrality class, shown with statistical uncertainties only.} 
    \label{fig:fig1}
    \end{center}
\end{figure}

Jet finding is performed using the \kt{} and anti-\kt{} algorithms implemented in FastJet \cite{fastjet}. Anti-\kt{} jets are used as signal jets of which \vj{} is reported. The median transverse momentum per area of \kt{} jets is used as an event-by-event estimate of \rhoav{}, the average energy density of the underlying event. Azimuthal anisotropy of the underlying event is modeled using the dominant \cite{flowpaper} flow harmonics $v_2$ and $v_3$  
\begin{equation}\label{eq:master}
    \rhovar = \rho_0 \left(1+2 \lbrace v_2 \cos\left[2 \left( \varphi - \ep \right)\right] +v_3 \cos\left[3\left(\varphi - \epthree \right)\right] \rbrace \right).
\end{equation}
Parameters $\rho_0$ and $v_n$ are determined event-by-event from a fit of the right side of Eq. \ref{eq:master} to the data; \rhovar{} is the azimuthal distribution of summed track \pt{} for tracks with 0.15 $<$ \pt{} $<$ 5 \GeVc{} and $\vert \eta \vert < $ 0.9. Figure \ref{fig:fig1} shows a single-event example. Measuring angles \epn{} in the V0 system removes short-range correlations between the event planes and tracks. The hard jet bias to flow harmonics $v_n$ is suppressed by rejecting all tracks around the jet axis for which $\vert \eta_{\rm jet} - \eta_{\rm track} \vert < R$. The corrected transverse jet momentum \ptjet{} is obtained by subtracting the local underlying event energy \rholocal, multiplied by the jet area $A$, from the raw jet momentum, $\ptjet = \ptraw - \rholocal \, A $, where  \rholocal{} is obtained from integration of \rhovar{} around $\phijet \pm R$, 
\begin{equation}\label{eq:int}
    \rholocal = \frac{\rhoav}{2 R \rho_0} \int_{\varphi - R}^{\varphi + R} \rhovar d \varphi.
\end{equation}
The pre-factor of the integral, $\frac{\rhoav}{2 R \rho_0}$, is chosen such that integration over the full azimuth yields the average energy density $\rhoav$.

The coefficient \vj{} is calculated from the difference between the unfolded \pt-differential jet yields in-plane (\nin) and out-of-plane (\nout),
\begin{equation}\label{eq:jetflow}
    \vj(\ptjet) = \frac{\pi}{4}  \frac{1}{\mathcal{R}_2} \frac{\nin(\ptjet) - \nout(\ptjet) }{\nin(\ptjet) +\nout(\ptjet) }.
\end{equation}
Equation \ref{eq:jetflow} is derived by integrating Eq.~\ref{eq:flowdud} for $n = 2$, over intervals $\left[ -\frac{\pi}{4}, \frac{\pi}{4} \right]$ and  $ \left[ \frac{3 \pi}{4}, \frac{5\pi}{4} \right]$ for \nin~and $\left[ \frac{\pi}{4}, \frac{3\pi}{4}\right]$ and $\left[\frac{5 \pi}{4}, \frac{7\pi}{4}\right]$ for \nout, substituting \ep{} for \rptwo{}. The factor $\mathcal{R}_2 \equiv \left< \cos \left[ 2 \left( \ep - \rptwo \right) \right] \right>$ corrects \vj{}  for the finite precision with which \ep{} approximates \rptwo{}. To correct the jet spectra for instrumental resolution and regional fluctuations in the energy density of the underlying event, \nin{} and \nout{} are unfolded independently, using independent descriptions of background energy-density fluctuations.

Systematic uncertainties on \vj{} are classified into two categories according to their point-to-point correlation. Shape uncertainties, dominated by unfolding, are anti-correlated between parts of the unfolded spectrum. Correlated uncertainties, mainly comprising tracking efficiency effects, are correlated point-to-point. Correlations between changes in \nin{} and \nout{} are taken into account for both categories.

\begin{figure}
    \includegraphics[width=.5\textwidth]{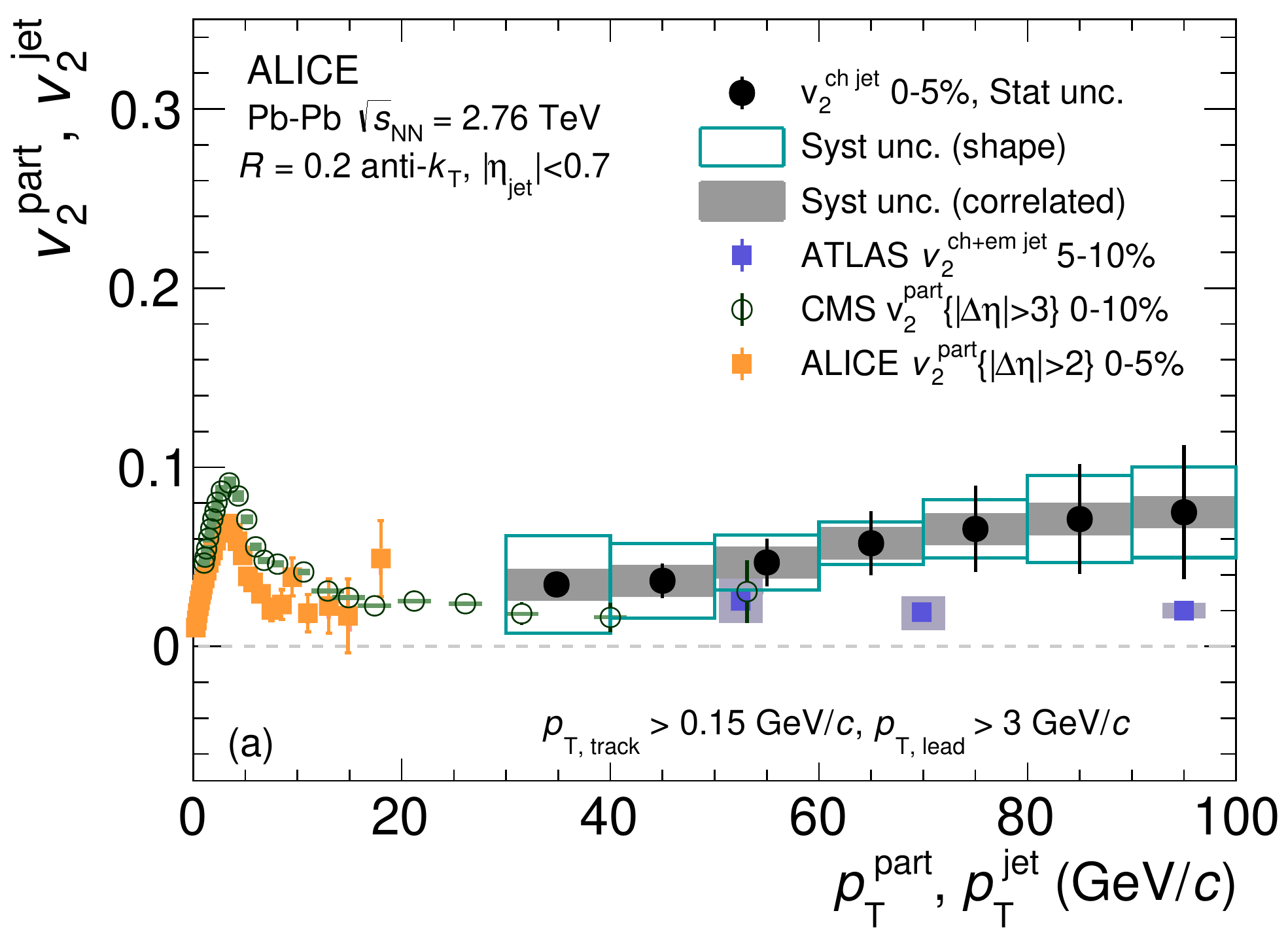}
    \includegraphics[width=.5\textwidth]{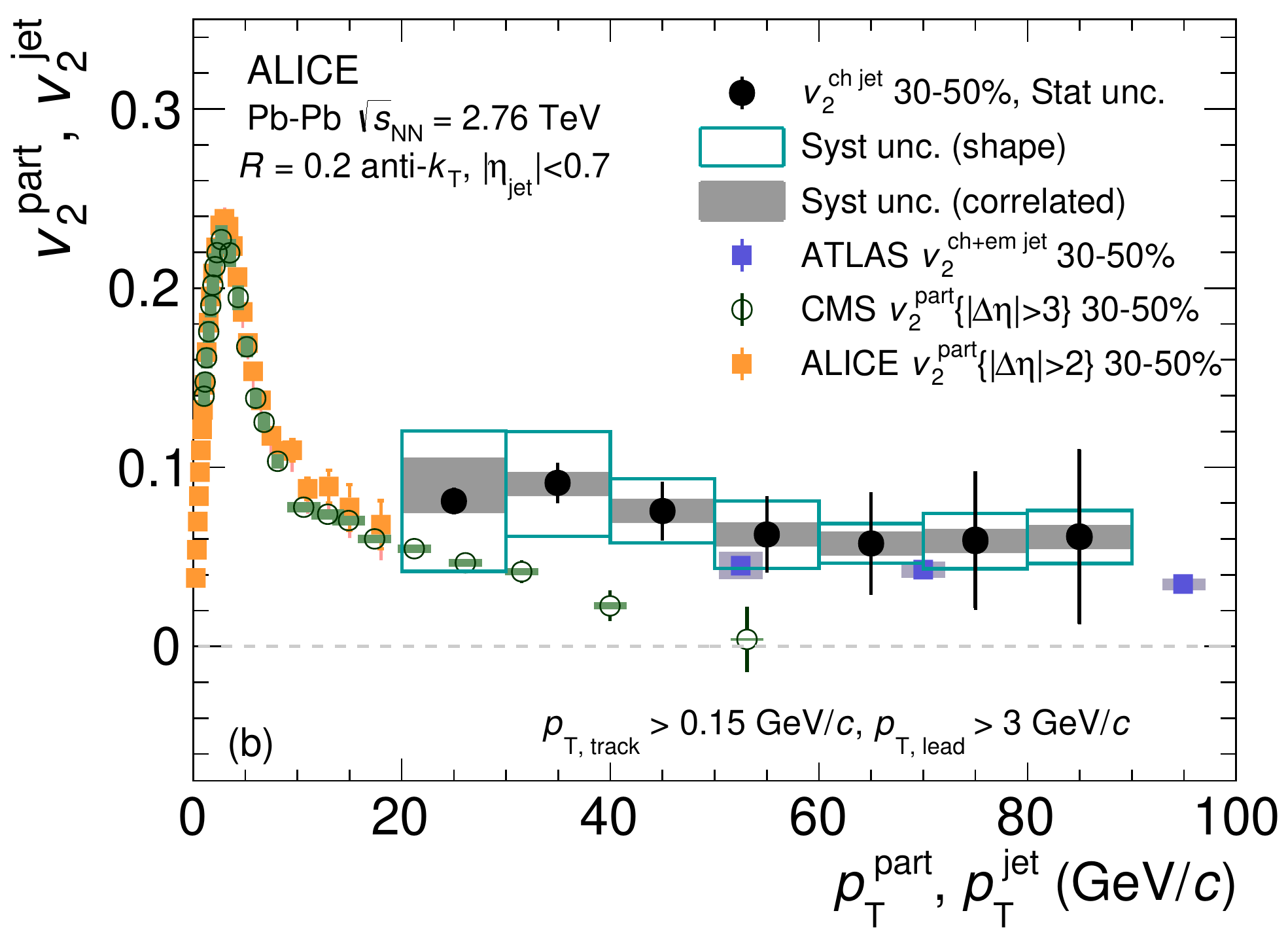}
    \caption{\vj{} of $R = 0.2$ charged jets as well as $v_2$ of charged particles \cite{highpt, highptv2CMS} (yellow, green) and \vjfull{} of $R = 0.2$ full jets (comprising both charged and neutral fragments) \cite{atlas} (blue). Statistical errors are represented by bars, systematic uncertainties by shaded or open boxes. Equal parton \pt{} corresponds to different single particle, full jet and charged jet \pt{}. ATLAS  \vjfull and CMS $v_2$ from \cite{highptv2CMS,atlas} in 30--50 \% centrality are the weighted arithmetic means of measurements in 10\% centrality intervals using the inverse square of statistical uncertainties as weights.  } 
    \label{fig:pel}
\end{figure}

\section{Results and comparisons}\label{sec:3}
Figure \ref{fig:pel} shows \vj{} in central (0-5\%) and semi-central (30-50\%) collisions. Significant positive \vj{} is found in semi-central collisions, indicating path-length-dependent in-medium parton energy loss. In central collisions, the larger relative contributions of background to the measured jet energy leads to larger systematic uncertainties. The compatibility of the data with a hypothesis of \vj{} = 0 is tested using a modified $\chi^2$ calculation as proposed in \cite{phenix} and is found to be within 1-2 standard deviations for the central and  3-4 standard deviations for the semi-central data. The \vpart{} of single charged particles \cite{highpt,highptv2CMS} and the ATLAS \vjfull{} measurement \cite{atlas} of $R = 0.2$ jets reconstructed from both charged and neutral fragments are superimposed in the same figure.
The central ATLAS results are reported in 5--10\% collision centrality. Qualitative agreement between the measurements is found, indicating path-length-dependent parton energy loss that is sensitive to the collision geometry up to high \pt{}. A quantitative comparison between the observables however cannot be made as equal parton \pt{} corresponds to different single particle, calorimeter jet and charged jet \pt{}.

\begin{figure}
    \includegraphics[width=.5\textwidth]{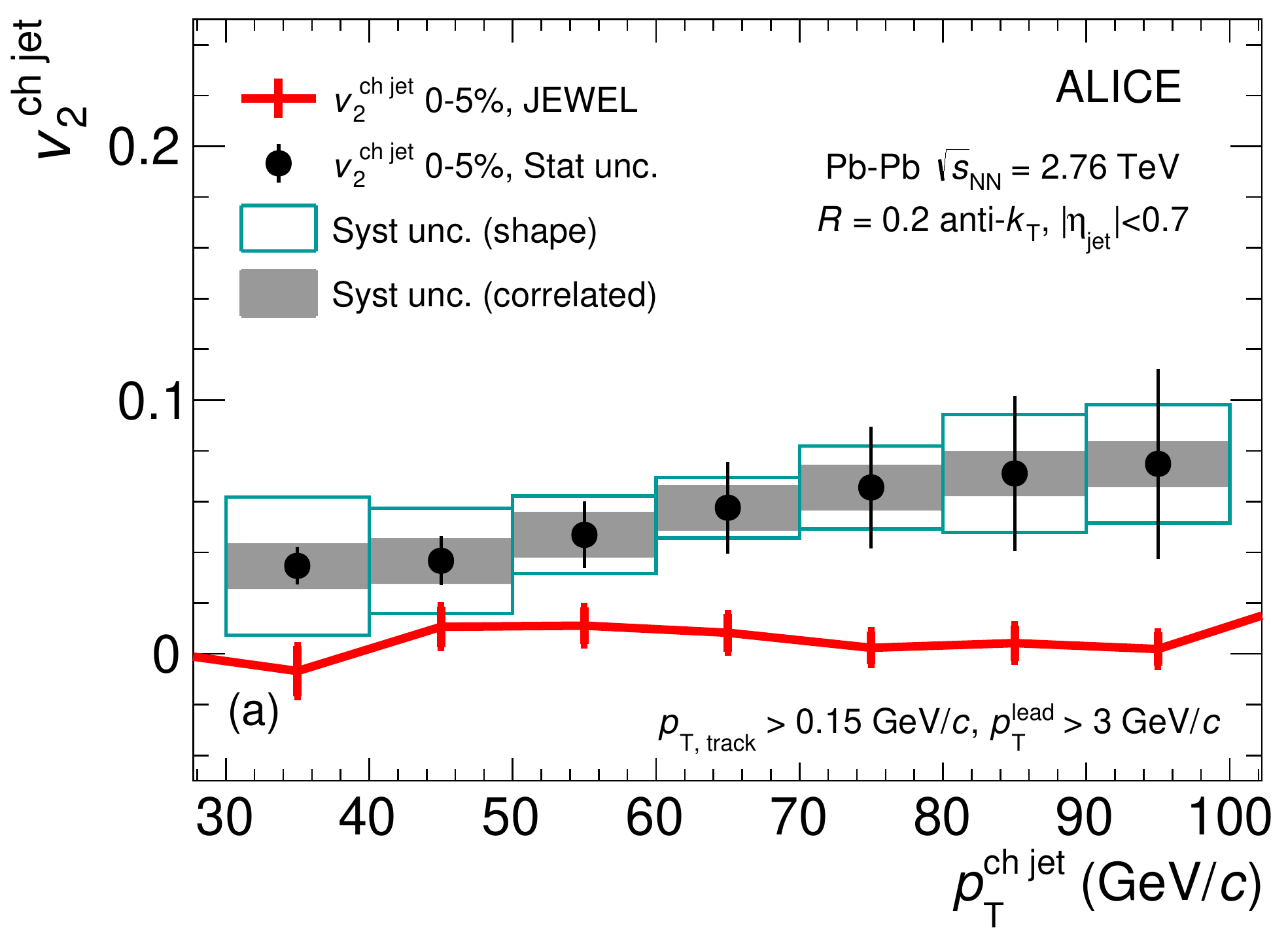}
    \includegraphics[width=.5\textwidth]{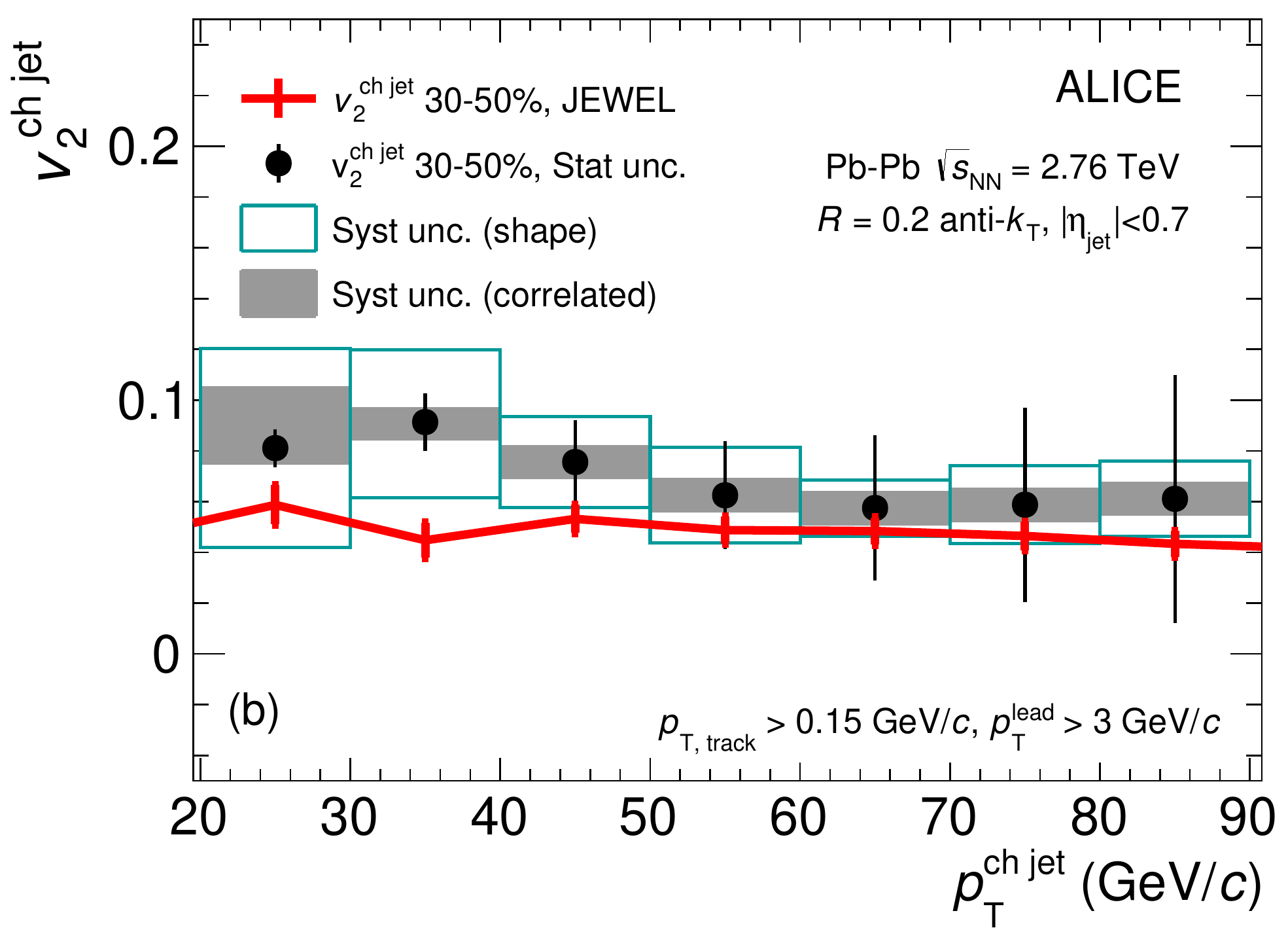}
    \caption{\vj~of $R = 0.2$ charged jets obtained from the JEWEL Monte Carlo (red) for central (a) and semi-central collisions (b) compared to data. JEWEL data points are presented with only statistical uncertainties. 
}
    \label{fig:jewel}
    \end{figure}

Figure \ref{fig:jewel} compares the data to \vj{} of $R = 0.2$ charged jets from the JEWEL Monte Carlo \cite{jewel1}, which models energy loss in the presence of a QCD medium through multiple scattering and gluon radiation. In semi-central collisions, good agreement with the model is found. JEWEL however currently does not include fluctuations in the participant distribution of nuclei in the nucleus, which may lead \cite{Ollitrault:1992bk,Alver:2006wh} to the apparent underestimation of \vj{} in central collisions. 

\section{Conclusion}
Measurements of \vj{} in central and semi-central Pb--Pb collisions have been presented. The data suggest that parton energy loss is large and that the sensitivity to the collision geometry persists up to high \pt{}. These findings are supported by measurements $v_2$ of single charged particles and full jets. The JEWEL Monte Carlo predicts sizable \vj{} for semi-central collisions and small to zero \vj{} in central events. The model is in good agreement with the semi-central measurement, indicating that energy loss can be described by scattering and gluon radiation. The comparison for central collisions suggests that initial state density fluctuations play an important role in generating \vj{}.

\bibliographystyle{elsarticle-num}
\bibliography{references}






\end{document}